# Point convolutional neural network algorithm for Ising model ground state research based on spring vibration

Zhelong Jiang[1,2], Gang Chen[1,*], Ruixiu Qiao[1], Pengcheng Feng[1,2], Yihao Chen[1,2], Junjia Su[1,2], Zhiyuan Zhao[1,3], Min Jin[1], Xu Chen[1], Zhigang Li[1], Huaxiang Lu[1,2,4,5]

[1]Institute of semiconductors, Chinese Academy of Sciences, Beijing, China

[2]Materials and Optoelectronics Research Center, University of Chinese Academy of Sciences, Beijing, China

[3]School of Microelectronics, University of Science and Technology of China, Hefei, China

[4]College of Microelectronics, University of Chinese Academy of Sciences, Beijing, China

[5]Semiconductor Neural Network Intelligent Perception and Computing Technology Beijing Key Laboratory, Beijing, China

[*]Email: chengang08@semi.ac.cn

## Abstract

The ground state search of the Ising model can be used to solve many combinatorial optimization problems. Under the current computer architecture, an Ising ground state search algorithm suitable for hardware computing is necessary for solving practical problems. Inspired by the potential energy conversion of springs, we propose a point convolutional neural network algorithm for ground state search based on spring vibration model, called Spring-Ising Algorithm. Spring-Ising Algorithm regards the spin as a moving mass point connected to a spring and establish the equation of motion for all spins. Spring-Ising Algorithm can be mapped on the GPU or AI chips through the basic structure of the neural network for fast and efficient parallel computing. The algorithm has very productive results for solving the Ising model and has been test in the recognized test benchmark $K_{2000}$. The algorithm introduces the concept of dynamic equilibrium to achieve a more detailed local search by dynamically adjusting the weight of the Ising model in the spring oscillation model. Finally, there is the simple hardware test speed evaluation. Spring-Ising Algorithm can provide the possibility to calculate the Ising model on a chip which focuses on accelerating neural network calculations.

## Introduction

Combinatorial optimization problems, a subfield of optimization with discrete variables, are ubiquitous in many fields of research. In many cases, we can find a mapping to the decision form of the Ising model with a polynomial number of steps for the NPC (Non-deterministic Polynomial Complete) problem [1], [2], [3], [4]. Therefore, many optimization problems can be formulated as Ising models to find the ground state, or the lowest energy configuration. So that, solving the Ising model becomes a general method for solving many NP problems, like partitioning problems [2], linear programming [1], [3], [5], inequality problems [6], coloring problems [2], [7] and so on. However, the Ising model is known to be NP-hard (Non-deterministic Polynomial Hard) problem [8]. So, it is difficult but important to find the ground state of the Ising model quickly and accurately.

The Ising model is mainly used in statistical physics and scientific computing. In statistical physics, the Ising model is widely used to study the phase transition phenomenon [9], [10], [11]. In scientific computing, the actual combinatorial optimization problem is mapped to the Ising model for finding the ground state in the N spins state space [12], [13], [14]. With N spins, there are $2^N$ spin state to search the global minimum of the energy state, which is a great challenge for using conventional computing [15]. Special-purpose hardware devices for the ground state search, known as Ising machines, have recently attracted attention because of their potential to substantially speed up the solution of optimization problems [16]. Various schemes have been proposed and demonstrated for the Ising model, including quantum annealers [17], [18], [19], [20], [21], coherent Ising machine [22], [23], [24], [25], [26], [27], [28], [29], [30], [31] and so on. Limited by current technology, the above methods have difficulties such as large-scale expansion and complicated parameter configuration. Quantum computer is expected to solve exponential combinatorial optimization problems, but related work is still in its infancy [32].

The CMOS implementations [16], [33], [34], [35], [36], [37] are easy to integrate and expand, which means that it is a more suitable strategy for mapping and solving large-scale practical Ising model problem. In application, CMOS Ising machine has advantages such as tiny size, flexible expansion, high integration, low system power consumption and so on [36]. Most CMOS chips are based on non-fully connected structures, like lattice graph [15], [33], [35], [36], king graph[34], [38], [39], [40], and Hexagonally graph [41]. All-to-all connected Ising models are more practical value than sparse ones, but communication and synchronization between the spins degrade the speed performance in CMOS [16]. As the result of that, the spin scale of a CMOS chip based on an all-to-all connected topology design is very limited. The non-uniform design limits the popularization of CMOS chips and increases the design cost of ASIC for Ising model.

AI (Artificial Intelligence) chips have numerous computing resources, which are used for training and Inference of various AI algorithms. and are important available resources for computing large-scale problems. At present, AI chips have solved many problems such as classification, detection, and tracking by virtue of their powerful computing power[42], [43]. Commercial AI chips have the characteristics of high energy efficiency, high parallelism, and high scalability. These chips, which are optimized for communication and synchronization, have been used in a large number of large models. The computing architecture of the AI chip sets the computing engine for the multiply accumulate operation and realizes the parallel computing through efficient scheduling, thereby reducing computing time and off-chip storage access [44]. Using these computing hardware resources to solve the Ising model with numerous parameters is an extremely effective method since we have not yet achieved quantum computers.

The paper is organized as follows. In this paper, we propose a new algorithm, Spring-Ising Algorithm, that can solve the all-to-all connected Ising model directly on the AI chip.

Frist, we introduce how Spring-Ising Algorithm inspired by spring vibrations can be used to find the ground state of the Ising model. Then, we design the algorithm as a network structure based on point convolution and residual modules, which implements the solution iteration of the Ising model through point convolution and residual modules. Through our method, the optimization problem is transformed into the general formula of AI chips calculation by constructing the Ising model paradigm and AI chips accelerate Spring-Ising Algorithm for the ground state finding. Finally, the network structure is demonstrated on AI chip architecture from Ref. [45] to solve the Max-cut problem and both numerical and analytical investigation are conducted.

# Result

In this chapter, we propose the physical prototype of Spring-Ising Algorithm and how to apply Lagrange's equations to iterate spin states by symplectic method. Spring-Ising Algorithm is inspired by physical phenomena, spring vibrations. The detail of physical prototype is introduced as follow.

1. Spring vibration model

The Ising model is defined as follow:

$$H = -\sum_{1\leq i<j\leq N} J_{ij}\sigma_i\sigma_j - \sum_{1\leq i\leq N} h_i\sigma_i \qquad (1)$$

The discrete variable $\sigma_i$ is the $i$th Ising spin state such that $\sigma_j \in \{-1,+1\}$. In Pauli matrices, the variable $\sigma_i$ assigns values $\{-1,+1\}$ to spin values $\{\downarrow,\uparrow\}$ [17]. $J_{ij}$ denotes a coupling coefficient between the $i$th and $j$th spins and $h_i$ is an external magnetic coefficient for the $i$th spin. $H$ is the total energy of the Ising model and it tends to the lowest energy.

Inspired by the steady-state analysis of multiple mass-spring system in analytical mechanics, the ground state research method of the Ising model in this paper is designed. In Ising model, the state of the $i$th spin ↑ (↓) is encoded as a discrete variable corresponding to a value of $+1(-1)$. We regard the discrete variable as the continuous change of the mass point in the macroscopic position, which is defined as the generalized coordinate $q_i \in [-1,1]$. On this basis, the spring model is designed by considering a mass point connected at an ideal spring with no initial length and the spring force on the mass point is always pointing to one point, called the origin point. As shown in Fig. 1(a), the spring is fixed at the origin point, and the other end is the mass point representing the state of spin. Since the initial length of the spring is zero, when the mass point moves away from the origin, it is pulled by the spring. In this model, the mass point is above(below) the origin to represent the spin ↑ (↓), and the distance is represented as a degree of confidence. According to the coupling coefficient and spin state, the Ising model produces a number of forces along a line along the $q_i$ axis. Therefore, the direction of the resultant force is also on the $q_i$ axis, as shown in Fig. 1(b).

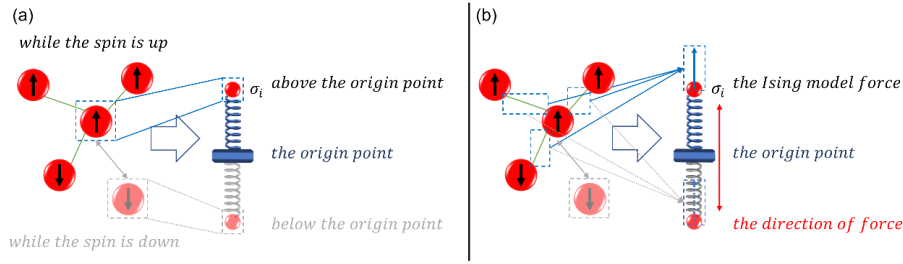

Fig 1. Spring vibration model based on Ising model. The red sphere represents the spin, and the arrow in it indicates the spin state. The four bright red spheres on the upper left represent the four spins mapped by the Ising model. The green connection line between the red spheres represents the coupling relationship. The fuzzy sphere in the gray dashed box represents the opposite spin state of the blue dashed box. The two dashed boxes are used to represent the same spin in two spin states, expressing the two particle positions of the spring model. Correspondingly, the gray part in the spring model is another spin state. (a) The spin of Ising model is mapping to the position of the mass point on the spring vibration model. Take the part in the blue dashed box as an example, the spin state is up, the mass point is above the origin. The blue dashed box vice versa. (b) The distance between the mass point and the origin point is effected by the coupling relationship and the spring.

In the model, while a spin considered as a mass point is called the target spin, the other spins are called the source spins providing external force to the target spins. The magnitude and direction of $F_i$ depend on the combined effect of multiple source spins but have nothing to do with the state of the target spin. Fig. 2(a) gives a specific example, when the state of source spin is $+1$, if the coupling coefficient is positive, an upward force will be generated. The greater the coupling coefficient, the greater the force generated. In the same way, if the coupling coefficient is negative, a downward force will be generated. When the coupling coefficient is zero, the source spin provides no force. The superposition of all the forces provided by the source spin is the force of the Ising model coupling relationship for the mass point $i$. When the state of origin spin is $-1$, the direction of the force is opposite, as shown in Fig. 2(b).

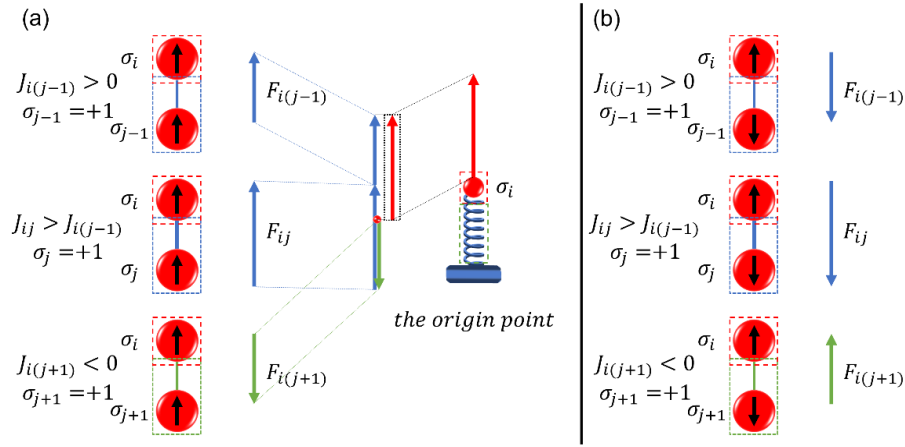

Fig 2. The specific example shows that the coupling relationship between spins affects the external force received on the mass point. $\sigma_i$ is the $i$th spin which is regarded as the target spin and $\sigma_j$ is the $j$th spin which is regarded as the source spin. The blue line between the spins means that the coupling relationship of the two spins is positive, the green line is negative. The force on the mass point is the resultant force produced by the sum of all coupling relations. (a) When the source spin $\sigma_i$ is $+1$, the coupling relationship produces multiple forces on the mass point $i$. (b) When the spin state $\sigma_i$ is $-1$, the direction of the force is opposite.

The generalized coordinate introduced by the model is a continuous variable, which means that the magnitude of the force is also affected by the absolute value of the generalized coordinate from the source spin. So, the source spin is changed to the source spin generalized coordinate: $\sigma_i \in \{-1,1\} \to q_i \in [-1,1]$. When the absolute value of the generalized coordinates is greater, the spring potential energy contained in the spring vibration model is greater. For the Ising model, the greater source spin has a greater overall influence on the system to the target spin and vice versa. Therefore, the discrete Ising model energy in Eq. (1) is set to the continuous Ising model energy in the spring vibration model.

2. Ground state search method

After establishing the spring vibration model, the following is how to use the model to find the ground state of Ising model. This method regards the potential energy of the Ising model as the ordinary potential energy and converts the potential energy of the Ising model into the potential energy of the spring and the kinetic energy of the system. The Ising model energy gradually decreases and transforms into the potential energy of the spring. Then, due to the constraints of generalized coordinates and generalized velocity, the energy originally converted into spring potential energy, continuous Ising model energy and kinetic energy is absorbed by the inelastic wall. The energy of the whole system gradually decreases and thus converges to the spring potential energy and the energy of the Ising model near the ground state.

According to the various constraints of this system, the Lagrangian equation is constructed as follows:

$$L(q_i, \dot{q}_i, t) = \sum_i m\dot{q}_i^2 - \sum_i \frac{1}{2}k(q_i - q_0)^2 - \zeta H_{Ising}(\boldsymbol{q}) \qquad (2)$$

Where $m$ is the mass coefficient, $k$ is the elastic coefficient, and $\zeta$ is the scaling coefficient of the Ising model energy. The three terms of the mass point in Eq. (2) are the kinetic energy term, the spring potential energy term and the continuous Ising model energy term. In the spring vibration model, the generalized coordinates are independent of $t$. It can be seen from the formula that the movement of the mass points is affected by the potential energy of the spring and the energy of the Ising model. The movement of the mass points is manifested as a continuous vibration on the ideal springs. From another perspective, it can be considered that when the spring is doing simple harmonic motion, a set of external forces are applied from the outside. Affected by the coupling coefficient of the Ising model, the oscillations of the mass points are biased towards the lower Ising model energy.

3. Symplectic method

Since the size of the Ising model depends on the number of spins, the solution scale is quite large. Therefore, it is very difficult to solve the Lagrangian equation directly and accurately. In this paper, referring to the Hamiltonian and symplectic method [46], the numerical iterative calculation of the spring vibration model is carried out. The Hamiltonian describes the total energy of the system and can be used to describe the system's dynamic behavior. Symplectic method is numerical method used to solve Hamilton's equations and it preserves energy conservation of the system.

According to the definition, the generalized momentum $p_i$ is obtained as $\partial L/\partial \dot{q} = m\dot{q}_i$. The Hamiltonian of the system is obtained by performing the Legendre transformation on the Lagrangian quantity:

$$H(q,p,t) = \sum_i \dot{q}_i p_i - L(q_i, \dot{q}_i, t) = \sum_i \frac{1}{2}\dot{q}_i p_i + \sum_i \frac{1}{2}k(q_i - q_0)^2 + \zeta H_{Ising}(\boldsymbol{q}) \qquad (3)$$

Get Hamilton's equation:

$$\begin{aligned}\dot{q}_i &= \frac{\partial H}{\partial p_i} \\ \dot{p}_i &= -\frac{\partial H}{\partial q_i} = -k(q_i - q_0) + \zeta \sum_j w_{ij} q_j\end{aligned} \qquad (4)$$

Use the symplectic algorithm to solve the Hamiltonian system, and set the system origin to $q_0$:

$$\begin{aligned} q_i(t_{n+1}) &= q_i(t_n) + \Delta \dot{q}_i(t_n) = q_i(t_n) + \frac{\Delta}{m} p_i(t_n) \\ p_i(t_{n+1}) &= p_i(t_n) + \Delta \dot{p}_i(t_n) = p_i(t_n) - \Delta k q_i(t_n) + \zeta \Delta \sum_j J_{ij} q_j(t_n) \end{aligned} \qquad (5)$$

Where $t_n$ is the $n$th iteration. It can be seen from the above formula that $q_i(t_n)$ and $p_i(t_n)$ depend on the value of the previous state. With the iteration of the value, the energy is continuously converted. As the energy of the Ising model decreases, the solution is gradually approaching the ground state of the Ising model. Dimensional issues are not considered in numerical calculations, so parameters can be combined. The Eq. (5) is called the iterative formula of Spring-Ising Algorithm.

4. Point convolutional neural network

In the iterative calculation of the algorithm, the most important calculation is multiplication of $J_{ij}$ and $q_i(t_n)$. A method of iterative calculation using point convolution to replace the product of vector and matrix is proposed, so that the algorithm can be used in high-bandwidth computing chips, like GPU and AI chip. Moreover, the point convolution is optimized in the hardware design to reuse the convolution kernel and reduce data access to accelerate the computation in parallel. Fig. 3 shows an alternative way of turning the iterative equation into the neural network architecture computation. $q_i(t_n)$ of a single test is assigned at fixed coordinate of the feature map, so the number of generalized coordinates and feature maps are the same. The size of the point convolution kernel also depends on the coupling relationship matrix of the Ising model so that the number of convolution kernel is $n$, and the number of convolution kernel channels is $n$. The rest of the architecture is the addition module, which can be completed through the residual structure in the neural network and is supported in mainstream AI chips.

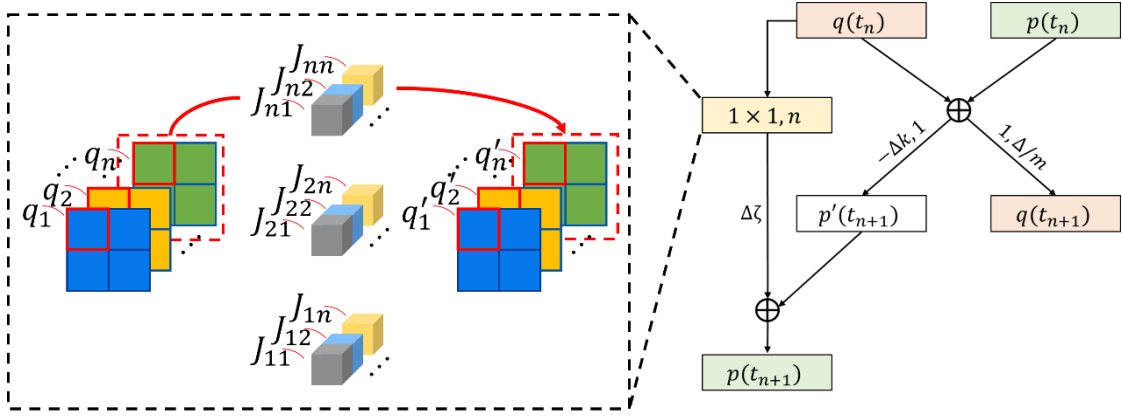

Fig 3. The parallel calculation of the spring vibration model algorithm through the form of point convolution. The size of the feature map affects the number of parallel tests for the algorithm. Using a 2×2 feature map is four independent iterative calculations. The value of the feature map is the generalized coordinate value, and the point convolution kernel is the weight data of the Ising model. The $q'_n$ is the temporary variable. On the right is the entire point convolution network architecture.

# Discussion

In this chapter, we show the experimental results based on the spring vibration model. Then we introduce how to implement the above algorithm through point convolution and residual network and implement it on the CASSANN-v2 architecture.

To demonstrate the effect of Eq. (1), the algorithm is tested on the $K_{2000}$ benchmark instance, which is a random undirected graph with 2000 vertices and 1999000 edges [23]. The $K_{2000}$ has been used many times to solve maximum-cut problems (MAX-CUT) for Ising model comparison of performance [23], [47], [48].

1. Qualitative result

When the kinetic energy is large enough, it can cross the local minimum value by the local oscillation; but at the same time, if the kinetic energy is too large, it cannot stay in any minimum value. So that, the following constraints are added each time $q_i$ is updated:

$$q_i \leftarrow f(q_i) = \begin{cases} -\sqrt{2}, & q_i < -\sqrt{2} \\ q_i, & -\sqrt{2} \leq q_i \leq \sqrt{2} \\ \sqrt{2}, & q_i > \sqrt{2} \end{cases} \quad (6)$$

Where $f(*)$ describes the boundaries of $q_i$. For the spring to vibrate, the boundary is slightly than the original setting $[-1,1]$ so that set $q_i \in [-\sqrt{2}, \sqrt{2}]$. After combining the boundary conditions, the equation describes the motion law of the spin.

The mass point vibration result of running the spring vibration model algorithm in 10000 iterations is shown in Fig. 4. The benchmark has 2000 spins, of which the first twenty are selected in the figure for visualization. In the early stage of the algorithm, because the mass points are initialized at origin and only given a small disturbance, the energy of the Ising model has a long-term decline process. It can be clearly seen in the figure that, the polylines are very dense, which means that the mass points are oscillating violently. In this time, the energy of Ising model is also rapidly oscillating and declining. In the middle, there are many mass points attached to the boundary gradually which have been at the lower energy point. Finally, only some mass points are still oscillating to search for the optimal and the energy of the Ising model has tended to the ground state and the detail of the energy change is shown in the inset of Fig. 4(a). Also, it can be seen that a few spin flips bring fluctuations of Ising energy.

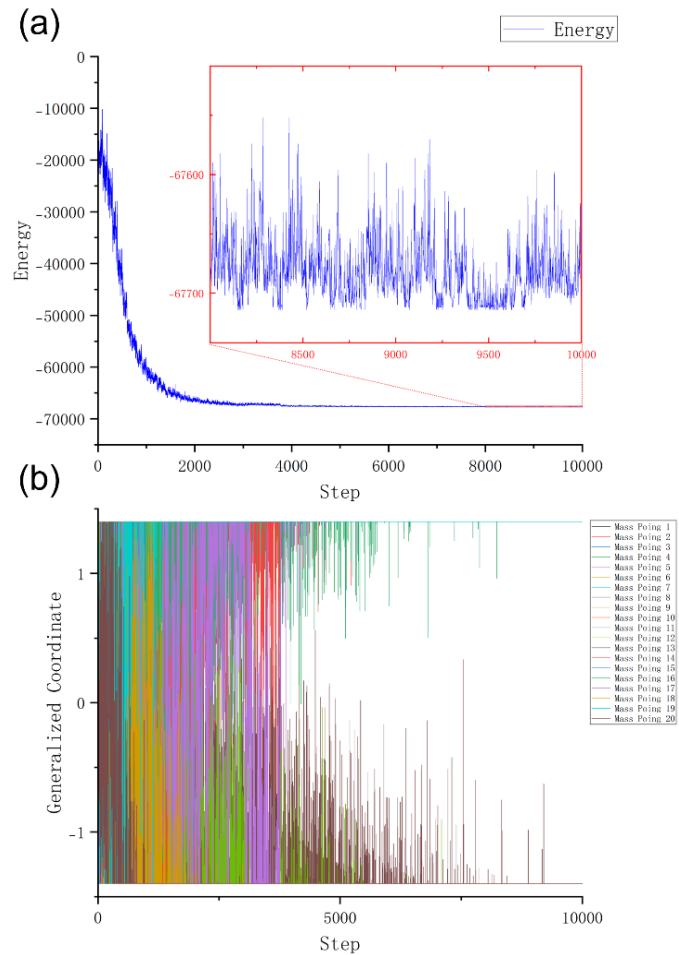

Fig 4. The spring vibration model algorithm on the $K_{2000}$ in 10000 iterations. The parameter configuration is as follows: k = 0.5, $\zeta = 0.8\zeta_0$ -> $10\zeta_0$, $\Delta$ = 0.2, m = 1. (a) The energy change curve of the Ising model. The mass point positions in Spring-Ising Algorithm are initialized near the origin, so the energy starts from 0 and decreases rapidly. Before $N_{step}$=2000, the energy is descending in a violent shock. After that, vibrate slightly to search for the energy minimum. (b) The mass point (the first twenty) vibration. The very dense parts of the graph are the effect of multiple mass points oscillations. When the system completes the basic search, it tends to be stable. Most of the mass points tend to be stable and only some of them perform local search (like after $N_{step}$=5000).

2. Quantitative result

It can be easily predicted that the potential energy of the spring is lost within the limitation of the boundary conditions as time progresses. Therefore, in the later stages of evolution, it is necessary to compensate for the lost energy. In order to further search the ground state of Ising model accurately, Spring-Ising Algorithm introduces the concept of energy dynamic balance to increase the energy proportion of Ising model and improve the search efficiency. To compensate for the energy loss, Spring-Ising Algorithm sets the $\zeta$ as a linear variable $\zeta(t)$. In order to reduce the complexity of the algorithm, this variable is regarded as a constant in the calculation of the Lagrangian equation, which means that the time-varying effect in the Lagrange equation is not considered. Through further analysis and solution of this equation, the ground state finding of the Ising model system is obtained.

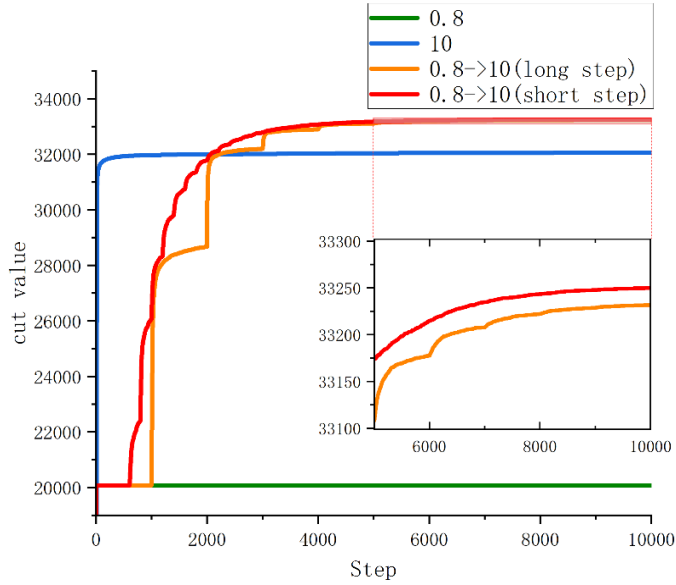

Fig 5. The effect of different $\zeta$ on the average results of $K_{2000}$. $\zeta_0$ is the base value. $\zeta_0$ = 0.05. The first and second sets of data (green curve and blue curve) indicate that the current $\zeta$ is fixed at $0.8\zeta_0$ or $10\zeta_0$, respectively. The third and fourth set of data (orange curve and red curve) indicates that the $\zeta$ is set from $0.8\zeta_0$ to $10\zeta_0$ with different step lengths. ($N_{step}$ = 200 or 1000)

This test is based on the same small disturbance for initializing with different strategies of $\zeta$. As shown in Fig. 5, no matter what the value of $\zeta$ is fixed, the ground state search of the Ising model is easy to fall into a local optimum. Although the larger $\zeta$ quickly leads to better local results (the blue line), it is difficult to search further to get better results. By gradually changing the value of $\zeta$, further searches can be performed after the spring model has entered local stability. The red line and the orange line can be clearly seen each time steady state is established and further searches. This result is very similar to sufficiently slow cooling in simulated annealing. When the step length is short enough, better search results can be obtained.

The probability density function (CDF) is an important way to judge the performance of algorithms for solving Ising models. Fig 6. shows the cumulative distributions of the cut value of the $K_{2000}$. The algorithm is compared with the HdSB and HbSB algorithm [47], and there are partial similarities under different modeling methods. It shows that the spring vibration model algorithm within the specified number of steps can search for better cut value. The inset shows that the algorithm can find the optimal value more effectively and the number of optimal solutions accounts for 2.9% of all solutions, which is only about 1.2% solved by HbSB and HdSB.

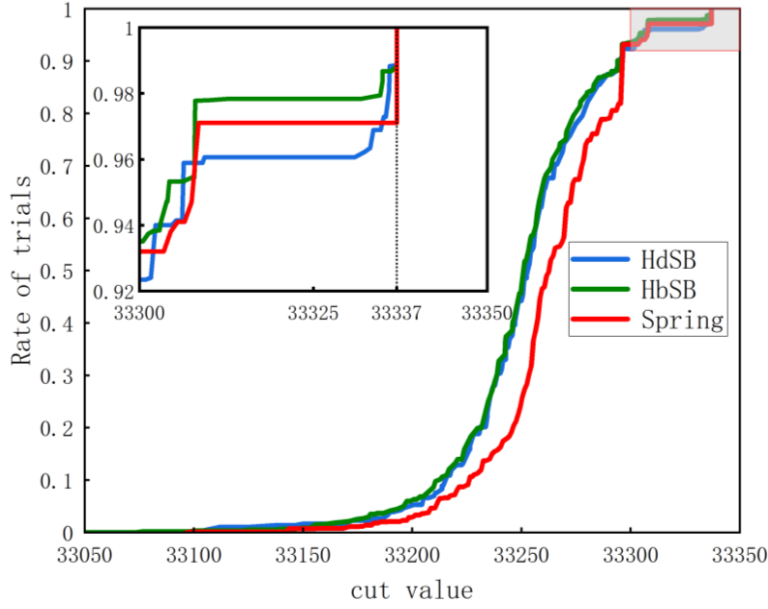

Fig 6. The spring vibration model algorithm cumulative distribution of cut values C of the $K_{2000}$ compared to HdSB and HbSB. The red curve is the result of the Spring-Ising Algorithm. The inset is the magnification around the best-known cut value. The red curve illustrate that the Spring-Ising Algorithm has better suboptimal distribution results and more optimal values than HdSB and HbSB for the overall search results.

3. Hardware implementation

The test platform of this algorithm is a personal computer (Intel 8700K and NVIDIA GTX 2080ti) and the AI architecture (CNN accelerator) developed by Institute of Semiconductors, CAS [45]. By GTX 2080ti testing in the PyTorch framework, when the size of Ising model is 2000 and the number of tests is 1000, the calculation time is 9.95s/10000 steps, which means that a sample time of 10,000-step tests is 9.95ms. But when the number of tests is 100, the time is 2.30ms/10000 steps. The GPU has a shorter average single-sample test time at more massive tests. The average power consumption of the 2080Ti is 60.6W. By the AI architecture, the size of Ising model is 2000 and the number of tests is 49 (7×7 feature map), the calculation time is 381.15ms/10000 steps, which means that a sample time of 10,000-step tests is 7.78ms. The average power consumption of the CASSANN-v2 is lower than 10W.

## Methods

1. Numerical iteration

First, regard the spin of the Ising model as $q$ and the coupling coefficient weight as $J$. The ground state search process of the Ising model is solved under the mass points oscillation process. Based on the spring vibration model, the vibration equation combined with the Ising model is constructed and transformed into the following Hamiltonian.

$$H(q,p,t) = \sum_i \frac{1}{2}\dot{q}_i p_i + \sum_i \frac{1}{2}k(q_i - q_0)^2 + \zeta H_{Ising}(\boldsymbol{q}) \qquad (7)$$

According to the equations of motion, all mass points are initialized near the origin, and then all mass points are made to start moving under the joint action of the spring and the Ising model. According to the symplectic method, the positions of the mass points at each time step are found. The final spin state of the Ising model is obtained by tracking the mass points (numerical iteration).

$$q_i(t_{n+1}) = q_i(t_n) + \Delta p_i(t_n)$$
$$p_i(t_{n+1}) = p_i(t_n) - \Delta k q_i(t_n) + \zeta(t_{step})\Delta \sum_j J_{ij} q_j(t_n) \quad (8)$$

$\Delta$, $k$ and $\zeta$ are independent adjustable variables. Then, Eq. (8) has the following constraints, like the perfectly inelastic walls work [47], $q_i \in [-\sqrt{2}, \sqrt{2}]$, $p_i \in [-2,2]$. $\zeta(t_{step})$ is a function linearly related to the number of iterations. For simple calculation, $\zeta(t_{step})$ is set as a piecewise constant function. Eq. (8) iterates the specified number of times to get the result of the Ising model ground state.

2. Hardware implementation

For an Ising model with $n$ spins, generalized coordinates $q$ are mapped to feature maps. The number of feature map pixels is the number of simultaneous iterations. The coupling coefficient matrix of the Ising model is mapped to the point convolution kernel. Divide the $J$ into $n$ 1×1 convolution kernels with $n$ channels by row. Through the residual structure, the addition operation required in the algorithm is completed. By continuously calling this network structure (Fig. 3), the numerical calculation of $q$ and $p$ in the eq. (8) is updated. After an artificially set time step or calculation time, the $q$ is sampled, which is the current low energy state of the Ising model.

# Data availability

The data that support the findings of this study are available from the corresponding author upon reasonable request.

# Code availability

The code used in this work is available from the corresponding author upon reasonable request.

# Acknowledgements

This study was supported by the National Natural Science Foundation of China [U19A2080] and the Strategic Priority Research Program of the Chinese Academy of Sciences [XDA27040303, XDA18040400, XDB44000000].

# Author contributions

Z. J. conceived the idea and developed the code. Z. J. and R. Q. conducted the numerical work. M. J., X. C. and Z. L. provided the theoretical foundations. All the authors contributed to write the paper. G. C. and H. L. guided and coordinated all aspects of the work.

## Competing interests

The authors declare no competing interests.